\documentclass[12pt,draftcls,onecolumn]{IEEEtran}

\usepackage{amsmath}
\usepackage{amssymb}
\usepackage{cite}
\usepackage{pgfplots}
\usetikzlibrary{calc}

\usepackage[latin1]{inputenc}
\usepackage{verbatim}

\usepackage{amsmath, amssymb, amsfonts,booktabs}													%
\usepackage{cite} 																											%

\usepackage{graphicx}
\usepackage{caption}
\usepackage{subcaption}

\usepackage[section,verbose]{placeins} 																			%
\usepackage{tikz}																											%
\usetikzlibrary{arrows,automata,matrix,fit,positioning,calc}											%
\usepackage{pgfplots}																										%
\usepackage{verbatim}																									%

\title{Parameter Space Design of Repetitive Controllers for Satisfying a Robust Performance Requirement}

\author{\IEEEauthorblockN{Burak Demirel\IEEEauthorrefmark{1}, and Levent G\"{u}ven\c{c}\IEEEauthorrefmark{1}\IEEEauthorrefmark{2} }
\thanks{
				\IEEEauthorblockA{\IEEEauthorrefmark{1}The authors are with the MEKAR (Mechatronics Research) Labs and Automotive Control and Mechatronics Research Center, Department of Mechanical Engineering, Istanbul Technical University, Istanbul TR-34437, Turkey (e-mail: demirelbu@itu.edu.tr; guvencl@itu.edu.tr).}
				\IEEEauthorrefmark{2} Corresponding author.
				}
}
\IEEEoverridecommandlockouts

\begin{document}

\maketitle

\begin{abstract}
A parameter space procedure for designing chosen parameters of a repetitive controller to satisfy a robust performance criterion is presented. Using this method, low order robust repetitive controllers can be designed and implemented for plants that possibly include time delay, poles on the imaginary axis and discontinuous weights. A design and simulation study based on a high speed atomic force microscope position control example is utilized to illustrate the method presented in this paper.
\end{abstract}


\IEEEpeerreviewmaketitle

\section{Introduction}\label{sec:Introduction}

\IEEEPARstart{R}{epetitive} controllers are used to accurately track a periodic reference signal or to reject a periodic disturbance with a known period by introducing a highly frequency selective gain through a positive feedback loop which contains a time delay element as this is a generator of periodic signals. The delay time is equal to the known period of the repetitive reference (or disturbance) signal. Repetitive control system is a special type of servo-system but its basic structure is based on the Internal Model Principle of Francis and Wonham~\cite{FrW:75}.
Significant improvements in the tracking accuracy or disturbance rejection characteristics of systems subject to periodic exogenous
signals can be achieved using repetitive control. The idea of repetitive control was first created by Inoue et al~\cite{IIN:81} to replace
conventional motion control techniques in the control of a proton synchrotron magnet power supply. Until recently, it has been
widely utilized in many application areas including control of hard disc drives~\cite{FKK:04}, control of optical disc drives~\cite{MLC:98}, control of noncircular tuning~\cite{TsT:94}, trajectory control of industrial robot arms~\cite{OHN:87, SHK+:90}, motor speed control~\cite{KHT:90}, high precision rotational control~\cite{FHC:00}, control of material testing machine~\cite{ShS:93}, control of cold rolling process~\cite{GaS:96}, suppression of torque vibration in motors~\cite{THI+:02}, reduction of waveform distortion in PWM inverter or UPS~\cite{RPG:03, ZKX:03, ZhW:03} and accurate position control of piezoelectric actuators~\cite{CLC:02, CJC:99}.

The earlier papers in the literature have generally focused on the stability analysis in both continuous time~\cite{HYO+:88, SrS:91} and discrete time systems~\cite{TTC:89}. Tsao and Tomizuka~\cite{TsT:94} have analyzed the robust stability of repetitive control systems applied to plants with unstructured modeling error. In order to achieve a specified level of nominal performance, Srinivasan et al~\cite{SOJ:95} have utilized the
Nevanlinna-Pick interpolation method to modify repetitive controller design by means of optimizing a measure of stability
robustness. Peery and \"Ozbay have modified $\mathcal{H}_{\infty}$optimal design approach presented in~\cite{Ozb:93} and then applied the extension of this methodology based on Youla parameterization to repetitive control systems in~\cite{PeO:97}. Moon et al~\cite{MLC:98} have developed a robust design methodology for parametric uncertainty in interval plants under repetitive control. Similarly, Roh and Chung~\cite{RoC:95} have created a new synthesis method based on Kharitonov's theorem for repetitive control systems with uncertain parameters. Weiss et al~\cite{WeH:99, WZG+:04, ZLW+:06} have made a stability and robustness analysis for MIMO
repetitive control systems based on $\mathcal{H}_{\infty}$ control theory. $\mu$ analysis has been used for assessing stability and performance robustness of SISO continuous time repetitive control systems by G\"uven\c{c}~\cite{Guv:96}. $\mu$ synthesis has been applied to sampled data repetitive controller design by Li and Tsao~\cite{LiT:98}.

The repetitive controller design approach presented in this paper is a continuation of the work presented in Aksun~G\"uven\c{c} and
G\"uven\c{c}~\cite{GuG:06} on repetitive controller design based on mapping the nominal performance and robust stability frequency domain constraints to controller parameter space where a servo-hydraulic material testing machine application was used. This work, in contrast, treats the robust performance constraint. Additionally, the efficiency of methodology is illustrated by using a high speed AFM scanner application. Moreover, the repetitive controller design approach presented in this paper is significantly different from those of the abovementioned references including the application of $\mathcal{H}_{\infty}$ methods. The significant advantages of the approach here in comparison with $\mathcal{H}_{\infty}$ methods are: (i) the ease of visualization due to the graphical representation of the solution in the parameter space approach and the capability and ease of doing multi-objective optimization by simply intersecting solution regions for different objectives, (ii) the
determination of a solution region rather than one specific solution for the control system satisfying a frequency domain
constraints (this makes it easier to design non-fragile controllers as changes in controller parameters will not violate the chosen objectives so long as the parameters are within the solution region), (iii) the determination of controller parameters that guarantee robust performance, (iv) being able to treat plants with time delay and poles on the imaginary axis, (v) not having to use rational, continuous weights in the robust performance specifications, and (vi) obtaining fixed structure low order repetitive controller filters that are easily implementable. There
are also some shortcomings of the proposed design method in comparison to the methods that exist in the literature including $\mathcal{H}_{\infty}$ methods such as: i) the method can simultaneously accommodate the design of only two controller parameters due to
its graphical display of the solution region, ii) the method does not result in a single analytical solution and the methods used do
not look mathematically elegant as a constructive frequency-by-frequency design approach is used.

It is difficult to apply standard robust control methods like
$\mathcal{H}_{\infty}$ control to repetitive controller design for robust
performance as the repetitive control system is infinite
dimensional due to the presence of the inherent time delay in the
controller. Robust control methods such as $\mathcal{H}_{\infty}$ optimal
control have been extended to infinite dimensional systems and
applied to repetitive control (see~\cite{SOJ:95} and~\cite{PeO:97}, for example).
However, very high order weighting functions need to be used in
the robust controller synthesis. Consequently, the resulting
repetitive controller filters also have high order. Model order
reduction techniques are used to reduce the order of the
repetitive controller filters in an actual implementation. Some of
the most powerful characteristics of the proposed method are that
the weights used in the design do not need to be continuous
functions of frequency and that plants can have time delay and/or
poles on the imaginary axis because the computations are naturally
carried out only at the frequencies of interest. Secondly, the
choice of the frequency grid used is not a problematic issue for
the repetitive control design procedure presented here as the main
design frequencies are exactly known and are the fundamental
frequency of the periodic exogenous signal (reference or
disturbance) with the period $\tau_d$ and its harmonics. The
largest harmonic frequency considered is chosen to be close to the
bandwidth of the repetitive control system which is limited by the
bandwidth of the actuator used in the implementation. The method
presented here is for SISO systems; however, it can be used to
design controllers for MIMO systems where one loop at a time
design is possible.

\subsection*{Paper Organization:}

The organization of the rest of the paper is as follows: Section 2
gives some basic information on robust repetitive controller
design. In Section 3, the technique of mapping robust performance
frequency domain specifications into repetitive controller
parameter space is explained in detail. Then, a numerical example
of a high speed AFM scanner position control application is
utilized in order to demonstrate the effectiveness of the proposed
method in Section 4. The paper ends with conclusions in Section 5.

\section{Repetitive Control Basics}

Consider the repetitive control structure shown in Fig. 1 where
$G_{n}$ is the nominal model of the plant, $\Delta_{m}$ is the
normalized unstructured multiplicative model uncertainty, $W_{T}$
is the multiplicative uncertainty weighting function and
$\tau_{d}$ is the period of the periodic exogenous signal. $q(s)$
and $b(s)$ are filters used for tuning the repetitive controller.
Repetitive control systems can track periodic signals very
accurately and can reject periodic disturbances very
satisfactorily. This is due to the fact that the positive feedback
loop in Fig. 1 is a generator of periodic signals with period
$\tau_{d}$ for $q(s)=1$. A low pass filter with unity d.c. gain is
used for $q(s)$ for robustness of stability~\cite{HYO+:88, WeH:99}.

\begin{figure}\centering
    \includegraphics[angle=0,width=0.65\columnwidth]{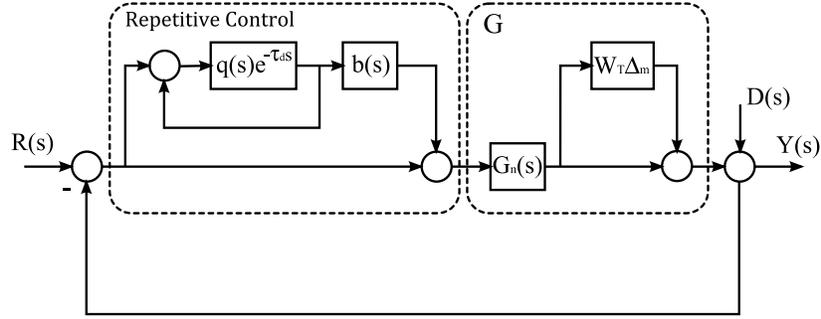}\\
    \caption{Repetitive control structure.}
\end{figure}

The repetitive controller design involves the design of the two
filters $q(s)$ and $b(s)$ seen in Fig. 1. In the frequency domain,
the ideal low-pass filter $q(j\omega)$ would be 1 in the frequency
range of interest and zero at higher frequencies. This is not
possible and $q(j\omega)$ will have negative phase angle which
will make $q(j\omega)$ differ from 1, resulting in reduced
accuracy. So as to improve the accuracy of the repetitive
controller, a small time advance is customarily incorporated into
$q(s)$ to cancel out the negative phase of its low-pass filter
part within its bandwidth. This small time advance can easily be
absorbed by the much larger time delay $\tau_{d}$ corresponding to
the period of the exogenous input signal and does not constitute
an implementation problem (see Fig. 2).

The main objective of the usage of the dynamic compensator $b(s)$ is improving the relative stability, the transition response and the steady state accuracy in combination with the low-pass filter $q(s)$. Consider the function of frequency given by
\begin{align}
	\mathcal{R}(\omega) \triangleq \bigg\vert q(j\omega) \Big[ 1-b(j\omega)\frac{G(j\omega)}{1+G(j\omega)} \Big] \bigg\vert \;,
	\label{eq:Regeneration_spectrum}
\end{align}
which is called the regeneration spectrum in~\cite{SrS:91}. $\mathcal{R}(\omega)<1-\epsilon$ for all $\omega\in [0,\infty )$ and some positive
$\epsilon$ is a sufficient condition for stability~\cite{SrS:91}. Moreover, $\mathcal{R}(\omega)$ can be utilized to obtain a good approximation of the locus of the dominant characteristic roots of the repetitive control system for large time delay, thus resulting in a measure
of relative stability, as well. Accordingly, the compensator $b(s)$ is designed to approximately invert $G/(1+G)$ within the
bandwidth of in an effort to minimize $\mathcal{R}(\omega)$. The dynamic compensator $b(s)$ can be selected as only a small time advance or
time advance multiplied by a low-pass filter in order to minimize $\mathcal{R}(\omega)$. In order to make $\mathcal{R}(\omega)<1$, the time advance in the filter $b(s)$ is chosen to cancel out the negative phase of $G/(1+G)$. This small time advance can easily be absorbed by the
much larger time delay $\tau_d$ corresponding to the period of the exogenous input signal and does not constitute an implementation
problem (see Fig. 2).

The $q(s)$ and $b(s)$ filters are thus expressed as
\begin{align}
	q(s)=q_{p}(s)e^{\tau_{q}s} \;,
\end{align}
and
\begin{align}
	b(s)=b_{p}(s)e^{\tau_{b}s} \;.
\end{align}

\begin{figure}\centering
    \includegraphics[angle=0,width=0.65\columnwidth]{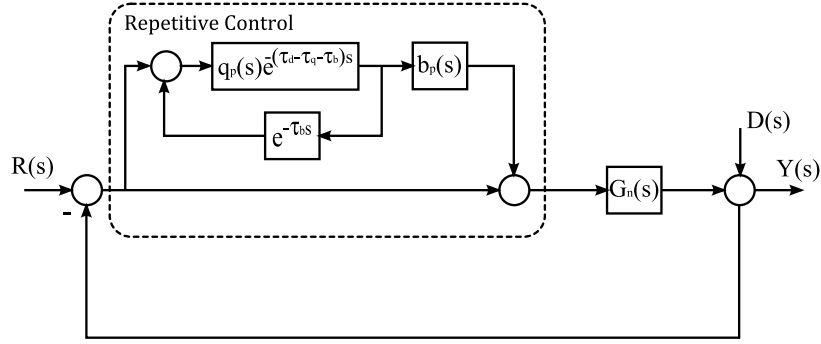}\\
    \caption{Modified repetitive control structure.}
\end{figure}

The time advances $\tau_{q}$ and $\tau_{b}$ are firstly chosen to decrease the magnitude of $\mathcal{R}(\omega)$ given in Eq.~\eqref{eq:Regeneration_spectrum}. Then, the
design focuses on pairs of chosen parameters in $q_{p}(s)$ or $b_{p}(s)$ to satisfy a frequency domain bound on the robust performance criterion. If $L$ denotes the loop gain of a control system, its sensitivity and complementary sensitivity transfer functions are
\begin{align}
	S \triangleq \frac{1}{1+L} \;,
\end{align}
and,
\begin{align}
	T \triangleq \frac{L}{1+L} \;.
\end{align}

The parameter space design, presented in the following section, aims at satisfying the condition
\begin{equation}
	\vert W_{S}S \vert + \vert W_{T}T \vert < 1\;, \qquad \forall\omega\in [0,\infty ) \;,
	\label{eq:Robust_performance_stability}
\end{equation}
which is similar to satisfying the robust performance requirement $\parallel\vert W_{S}S \vert + \vert W_{T}T \vert\parallel_{\infty}<1$ where $W_{S}$ and $W_{T}$ are sensitivity and complementary sensitivity function weights.

The loop gain of the repetitive control system seen in Figs. 1 and 2 is given by
\begin{align}
	L=G_{n}\Big(1+\frac{q_{p}}{1-q_{p}e^{(-\tau_d+\tau_q)s}}b_{p}e^{(-\tau_d+\tau_q+\tau_{b})s}\Big) \;.
\end{align}

The robust performance design requires
\begin{align}
	\vert W_{S}(\omega)S(j\omega) \vert + \vert W_{T}(\omega)T(j\omega) \vert =
	\bigg\vert \frac{W_{S}(\omega)}{1+L(j\omega)} \bigg\vert + \bigg\vert \frac{W_{T}(\omega)L(j\omega)}{1+L(j\omega)} \bigg\vert < 1 \;,
\quad \forall\omega\in [0,\infty ) \;,
\label{eq:Point_condition_v2}
\end{align}
or equivalently,
\begin{align}
	\vert W_{S}(\omega) \vert + \vert W_{T}(\omega)L(j\omega) \vert < \vert 1+L(j\omega) \vert \;, \quad \forall\omega\in [0,\infty )\;,
	\label{eq:Point_condition}
\end{align}
to be satisfied.

\section{Mapping Robust performance Frequency Domain Specifications into Repetitive Controller Parameter Space}

In the present section, a repetitive controller design procedure based on mapping the robust performance frequency domain performance specification given in Eq.~\eqref{eq:Point_condition} with an equality sign into the chosen repetitive controller parameter plane at a chosen frequency is described.

Consider the robust performance problem given in Fig. 3 illustrating Eq.~\eqref{eq:Point_condition} with an equality sign (called the robust performance point condition). Apply the cosine rule to the triangle with vertices at the origin, $-1$ and $L$ in Fig. 3 to obtain
\begin{align}
	( \vert W_{S}(\omega) \vert + \vert W_{T}(\omega)L(j\omega) \vert )^{2} = \vert L(j\omega) \vert^2 + 1^2 + 2\vert L(j\omega)\vert \cos{\theta_{L}} \;. \label{eq:QuadraticEquation}
\end{align}

Equation~\eqref{eq:QuadraticEquation} is a quadratic equation in $\vert L(j\omega) \vert$ and its solutions are
\begin{align}
	\vert L(j\omega) \vert = \frac{( -\cos{\theta_L} + \vert W_{S}(\omega) \vert \vert W_{T}(\omega) \vert )\pm\sqrt{\Delta_{M}}}{1 - \vert W_{T}(\omega) \vert^2} \;, \label{eq:Solution_QuadraticEquation}
\end{align}
where
\begin{align}
	\Delta_{M}=\cos^{2}{\theta_L} + \vert W_{S}(\omega) \vert^{2} + \vert W_{T}(\omega) \vert^{2} - 2\vert W_{S}(\omega) \vert^{2} \vert W_{T}(\omega) \vert^{2}\cos{\theta_{L}}-1 \;. \label{eq:Determinant}
\end{align}

Only, positive and real solutions for $\vert L \vert$ are allowed, i.e., $\Delta_{M}\geq0$ in Eq.~\eqref{eq:Solution_QuadraticEquation} must be satisfied. The point condition solution procedure is outlined below.

\textbf{M1.} Define the set of frequencies to be used as
\begin{align*}
	\Omega=\big\{\omega_{1},\omega_{2},\ldots,\omega_{n};\omega_{n+1},\omega_{n+2},\ldots,\omega_{m}; \omega_{m+1},\omega_{m+2},\ldots,\omega_{l} \big\},\omega_{k}=\frac{2\pi
k}{\tau_{d}}, k=1,2,\ldots,l 
\end{align*}
where $\omega_{1}=2\pi/\tau_{d}$ is the frequency of the periodic
exogenous input and  $\omega_{k}=2\pi k/\tau_{d}$ is the chosen
bandwidth of repetitive control (limited by the bandwidth of the
actuator used). Frequencies $\omega_{m+1}$ to $\omega_{l}$ are
high frequencies where significant model uncertainty exists
($\omega_{m+1}>10\omega_{n}$) and the intermediate frequencies
$\omega_{n+1}$ to $\omega_{m}$.

\emph{Remark:} It should be noted that the inherent time delay in
a repetitive control system will improve performance only at the
fundamental frequency $\omega_{1}=2\pi/\tau_{d}$ and its
harmonics. Repetitive control will worsen performance at
frequencies between the fundamental frequencies and harmonics. For
this reason, repetitive control is only used in the presence of a
periodic external input (reference or disturbance) as it will
result in degraded performance for non-periodic external inputs.
For that reason, the weights ($W_{S}$ and $W_{T}$) can be assumed
to be zero outside the finite set $\Omega$ given in step M1. Once
the design is complete, the designer checks the $\vert S \vert$ plot at low
frequencies and the $\vert T \vert$ plot at high frequencies to make sure
that the magnitude envelopes corresponding to intermediate
frequencies (between the harmonics) are at an acceptable level.
Another approach will be to specify weights for intermediate
frequencies in between the fundamental frequency and harmonics. A
major deficiency of parameter space methods is that one needs to
use a large number of frequencies in the set $\Omega$ to make sure
that the robust performance condition will not be violated at
frequencies outside of $\Omega$. In the case of repetitive
control, this problem is less severe at low frequencies where the
designer is interested mainly in reducing the sensitivity function
at the fundamental frequency and its harmonics. A large number of
frequency points can be used at higher frequencies above the
bandwidth of the repetitive controlled system.

\begin{figure}\centering
    \includegraphics[angle=0,width=0.4\columnwidth]{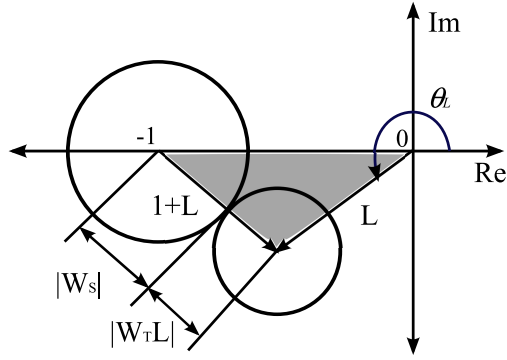}\\
    \caption{Illustration of the point condition for robust performance.}
\end{figure}

\textbf{M2.} Choose a specific frequency value $\omega=\omega_{i}\in\Omega$, $i=1,2,\ldots,l$ from set $\Omega$ in step M1. $\vert W_{S}(\omega) \vert$,$\vert W_{T}(\omega) \vert$ and $\vert G(\omega) \vert$ at a frequency $\omega$ are known at this point.

\textbf{M3.} Let $\theta_{L}\in[0,2\pi]$. Evaluate $\Delta_{M}$ by using Eq.~\eqref{eq:Determinant} and select the active range of $\theta_{L}$ where $\Delta_{M}\geq 0$ is satisfied. For all values of $\theta_{L}$ in the active range:

\textbf{M3a.} Evaluate $\vert L \vert$ by using Eq.~\eqref{eq:Solution_QuadraticEquation}. Keep only the positive solutions.

\textbf{M3b.} Evaluate $L = \vert L \vert e^{j\theta_{L}}$.

\textbf{M3c.} Solve for the corresponding repetitive controller filters $q_{p}(j\omega)$ and $b_{p}(j\omega)$ at the chosen frequency $\omega$ by utilizing
\begin{align}
	q_{p}(j\omega)=\frac{L(j\omega)-G(j\omega)}{L(j\omega)-G(j\omega)[1-b(j\omega)]}e^{(\tau_{d}-\tau_{q})j\omega} \;,
\end{align}
and
\begin{align}
	b_{p}(j\omega)=[L(j\omega)-G(j\omega)]\Big[\frac{1-q(j\omega)e^{-\tau_{d}j\omega}}{q(j\omega)e^{-\tau_{d}j\omega}}\Big]e^{-\tau_{b}j\omega} \;.
\end{align}

\textbf{M3d.} Using the specific structure of $q_{p}(j\omega)$ or $b_{p}(j\omega)$, back solve for the two chosen controller parameters within them. For example, $q_{p}(s)$ and $b_{p}(s)$ can be chosen as a multiplication of the second order controllers given by
\begin{align}
	q_{p}(s)=\prod_{i=1}^{n}\frac{q_{5i}s^{2}+q_{4i}s+q_{3i}}{q_{2i}s^{2}+q_{1i}s+q_{0i}}, \quad \text{and} \quad
~b_{p}(s)=\prod_{i=1}^{n}\frac{b_{5i}s^{2}+b_{4i}s+b_{3i}}{b_{2i}s^{2}+b_{1i}s+b_{0i}} \;. 
	\label{eq:Second_order_filters}
\end{align}
There are six tunable parameters for $n=1$ in Eq.~\eqref{eq:Second_order_filters} which can be used to represent different types of controllers. These six tunable parameters are $q_{5i}$, $q_{4i}$, $q_{3i}$, $q_{2i}$, $q_{1i}$ and $q_{0i}$ for $q_{p}(s)$. For $n=1$, the controller
structure in~\eqref{eq:Second_order_filters} consists of some well-known controller types such as proportional-integral-derivative (PID), lead-lag
controller, first or second order filters as illustrated in Table~1. If the performance of the filters which are utilized in the repetitive controller for $n=1$ is unsatisfactory, $n$ can be increased and new higher order filters can be synthesized. For the filter structure choice in Eq.~\eqref{eq:Second_order_filters}, the back solution procedure uses
\begin{align}
	\begin{array}{rcl}
		\mathrm{Re}[q_{p}(j\omega)]=\frac{(q_{3i}-q_{5i}\omega^{2})(q_{0i}-q_{2i}\omega^{2})+q_{1i}q_{4i}\omega^{2}}{(q_{0i}-q_{2i}\omega^{2})^{2}+(q_{1i}\omega)^{2}} \;, \\
		\mathrm{Im}[q_{p}(j\omega)]=\frac{q_{4i}\omega(q_{0i}-q_{2i}\omega^2)-q_{1i}\omega(q_{3i}-q_{5i}\omega^{2})}{(q_{0i}-q_{2i}\omega^{2})^{2}+(q_{1i}\omega)^{2}} \;,
	\end{array}
\end{align}
and
\begin{align}
	\begin{array}{rcl}
		\mathrm{Re}[b_{p}(j\omega)]=\frac{(b_{3i}-b_{5i}\omega^{2})(b_{0i}-b_{2i}\omega^{2})+b_{1i}b_{4i}\omega^{2}}{(b_{0i}-b_{2i}\omega^{2})^{2}+(b_{1i}\omega)^{2}} \;, \\
		\mathrm{Im}[b_{p}(j\omega)]=\frac{b_{4i}\omega(b_{0i}-b_{2i}\omega^2)-b_{1i}\omega(b_{3i}-b_{5i}\omega^{2})}{(b_{0i}-b_{2i}\omega^{2})^{2}+(b_{1i}\omega)^{2}} \;.
	\end{array}
\end{align}

\textbf{M4.} The solution in step M3 above results in a closed curve which is plotted for solving two of the twelve parameters
$q_{0i}$ to $q_{5i}$ and $b_{0i}$ to $b_{5i}$. Plot the closed curve obtained in the chosen controller parameter space. Either
the inside (drawn with a solid boundary) or outside (drawn with a dashed boundary) of this curve is a solution of Eq.~\eqref{eq:Point_condition_v2} at the chosen frequency (see the ellipses in Fig.~5, for example). The region obtained is the point condition solution in the chosen repetitive controller parameter plane at the frequency chosen in step M2.

\textbf{M5.} Go back to step M2 and repeat the procedure at a different frequency until all frequencies in set $\Omega$ are used.

\textbf{M6.} Plot the intersection of all point condition solutions for all frequencies in set $\Omega$. This is the overall solution region for the robust performance requirement.

As the solution procedure only uses frequency response data and is numerical in nature, plants with time delay or poles on the imaginary axis and discontinuous weights do not pose any problems. Note that solution regions for nominal performance $\vert W_{S}S \vert < 1$ for all $\omega\in [0,\infty )$ and for robust stability $\vert W_{T}T \vert < 1$ for all $\omega\in [0,\infty )$ can easily be obtained using the algorithm above
by setting $W_{S}=0$ and $W_{T}=0$, respectively. It is then possible to concentrate on nominal performance at low frequencies,
robust performance at intermediate frequencies and robust stability at high frequencies, obtaining three solution regions.
The overall solution region in the controller parameter space is then determined by the intersection of all three regions for
nominal performance, robust performance and robust stability. This procedure is easily programmable and quickly results in a
controller parameter space representation of the solution. The controller parameter space presentation obtained offers the ease
of visualization of parameter space methods (see Fig. 5) when one accepts the shortcoming of treating only two controller parameters
at a time. Multi-objective design can easily be formed in parameter space as it amounts simply to intersection of individual
solution regions. It is also possible to determine the final design (or tuning point) by optimizing some other criteria, such
as nominal time domain performance within the solution region obtained. In contrast to $\mathcal{H}_{\infty}$ optimal control synthesis,
there is no relationship between the order of repetitive control filters and the complexity of weights in this proposed method. The
main strength of this method is that low-order, easily implementable repetitive control filters are specified from the beginning.

\begin{table}[!t]
\renewcommand{\arraystretch}{1.3}
\caption{Controller Coefficients Table} \label{table_example}
\centering
\begin{tabular}{c||c||c||c||c||c||c||c}
\hline
\bfseries Control Action & \bfseries $n$ & \bfseries $q_{5i}$ & \bfseries $q_{4i}$ & \bfseries $q_{3i}$ & \bfseries $q_{2i}$ & \bfseries $q_{1i}$ & \bfseries $q_{0i}$\\
\hline\hline
P & 1 & 0 & 0 & $K$ & 0 & 0 & 1 \\
\hline
PD & 1 & 0 & $K T_{d}$ & $K$ & 0 & 0 & 1 \\
\hline
PI & 1 & 0 & $K$ & $K T_{i}$ & 0 & 1 & 0 \\
\hline
PID & 1 & $K T_{d}$ & $K$ & $K T_{i}$ & 0 & 1 & 0 \\
\hline
Lag ($\beta>1$) & 1 & 0 & $K T$ & $K$ & 0 & $\beta T$ & 1 \\
\hline
Lead ($0<\alpha<1$) & 1 & 0 & $K T$ & $K$ & 0 & $\alpha T$ & 1 \\
\hline
$1^{st}$ Order Filter & 1 & 0 & 0 & $K$ & 0 & $\tau$ & 1 \\
\hline
$2^{nd}$ Order Filter & 1 & 0 & 0 & $K \omega^2$ & 1 & $2\zeta\omega$ & $\omega^2$ \\
\hline
\end{tabular}
\end{table}

It is possible that for certain data sets $\vert W_{S} \vert$, $\vert W_{T} \vert$,
$G$, $\omega$; no solutions to the solution procedure outlined
above exist. Nonexistence of a solution for a specific frequency
$\omega$ could be because of nonexistence of a positive
$\Delta_{M}$ in Eq.~\eqref{eq:Determinant} or nonexistence of a positive solution
$\vert L \vert$ in Eq.~\eqref{eq:Solution_QuadraticEquation}. Nonexistence of a solution usually results from
a weight $\vert W_{S} \vert$ or $\vert W_{T} \vert$ that is too restrictive. The
solution procedure, which is programmed in an interactive fashion,
results in no solution points in this case. Then, the user will
know that his robust performance requirement at that frequency was
too restrictive and has the choice of relaxing this requirement.
Note that solutions might exist at all individual frequencies,
however; their intersection in Step M6 resulting in the overall
solution region, might still be empty. In that case, the user must
change the sensitivity and complementary sensitivity weights at
the problematic frequencies.

\section{Numerical Example}

In this part of the paper, the high speed atomic force microscope (AFM) scanner which is designed and modeled in~\cite{SAD+:07} is utilized as a numerical example to explain the methodology of the multi-objective parameter space approach for SISO repetitive controller design. The second order and fourth order mathematical models of this high speed AFM scanner are given in~\cite{SAD+:07}. In this example, the fourth order model is used because it includes the first mode of the piezoelectric stack in the vertical direction. The transfer function of the AFM scanner is given by
\begin{align}
	G(s)=\frac{K(s^2+2\zeta_{2}\omega_{2}s+\omega_{2}^2)}{(s^2+2\zeta_{1}\omega_{1}s+\omega_{1}^2)(s^2+2\zeta_{3}\omega_{3}s+\omega_{3}^2)} \;, \label{eq:AFM_MathematicalModel}
\end{align}
where $K=1\times10^{12}$nm/V includes the power amplifier and sensor gain. The system seen in~\eqref{eq:AFM_MathematicalModel} has two resonant frequencies and one anti-resonant frequency. The numerical values of these frequencies are given as $f_{1}=40.9$ kHz, $f_{2}=41.6$ kHz and $f_{3}=120$ kHz and can be seen in Fig. 4. The numerical values of the relative damping coefficients are given as $\zeta_{1}=0.016$, $\zeta_{2}=0.016$ and $\zeta_{3}=0.17$. The dynamic compensator $b(s)$ is chosen as a pure time advance as
\begin{align}
	b(s)=b_{p}(s)e^{\tau_{b}}=e^{3\times10^{-6}s} \;.
\end{align}

The low-pass filter $q(s)$ is chosen as
\begin{align}
	q(s)=q_p(s)e^{\tau_{q}s}=\frac{a_{0}}{s^2+a_{1}s+a_{0}}e^{7.5\times10^{-6}s} \;.
	\label{eq:Low_pass_filter}
\end{align}

The parameters of $q_{p}(s)$ given in Eq.~\eqref{eq:Low_pass_filter} are chosen as $q_{51}=q_{41}=0$, $q_{21}=1$, $q_{31}=q_{01}=a_{0}$ and
$q_{11}=a_{1}$ in the general form~\eqref{eq:Second_order_filters} in order to obtain unity d.c. gain. Phase advance is also added to this low-pass filter phase cancellation. Thus, a decrease in the steady state error is aimed. The region in the $q_{01}-q_{11}$ controller parameter
space are computed for three cases which are respectively the nominal performance at low frequencies ($W_{T}=0$), robust
performance at intermediate frequencies and robust stability at high frequencies ($W_{S}=0$).

\begin{figure}\centering
    \includegraphics[angle=0,width=0.5\columnwidth]{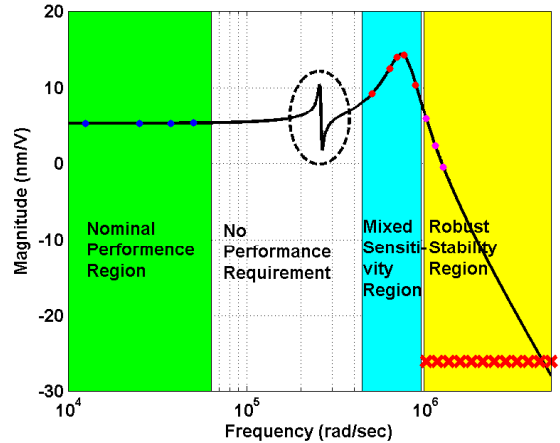}\\
    \caption{The Bode magnitude plot of high speed AFM-scanner with the mapping
    frequencies for the nominal performance, robust performance and robust stability.}
\end{figure}

\begin{table}[!t]
\renewcommand{\arraystretch}{1.3}
\caption{Desired Sensitivity Magnitude Upper Bounds at
$\tau_{d}=0.0005sec$} \label{table_example} \centering
\begin{tabular}{c||c||c||c||c}
\hline
\bfseries Frequency Range & \bfseries $k$ & \bfseries $f=k/\tau_{d}(kHz)$ & $W_{S}$ & $W_{T}$\\
\hline\hline
Low (NP) & 1 & 2 & 500 & 0\\
\hline
Low (NP) & 2 & 4 & 225 & 0\\
\hline
Low (NP) & 3 & 6 & 115 & 0\\
\hline
Low (NP) & 4 & 8 &  75 & 0\\
\hline
Intermediate (RP) & 40 &  80 & 3.3 & 0.001\\
\hline
Intermediate (RP) & 50 & 100 & 4.5 & 0.045\\
\hline
Intermediate (RP) & 55 & 110 & 4.5 & 0.001\\
\hline
Intermediate (RP) & 60 & 120 & 1.5 & 0.005\\
\hline
Intermediate (RP) & 70 & 140 & 1.5 &  0.01\\
\hline
High (RS) &  80 & 160 & 0 & 0.05\\
\hline
High (RS) &  90 & 180 & 0 & 0.05\\
\hline
High (RS) & 100 & 200 & 0 & 0.05\\
\hline
\end{tabular}
\end{table}

\begin{figure}
        \centering
        \begin{subfigure}[b]{0.5\textwidth}
                \includegraphics[width=\textwidth]{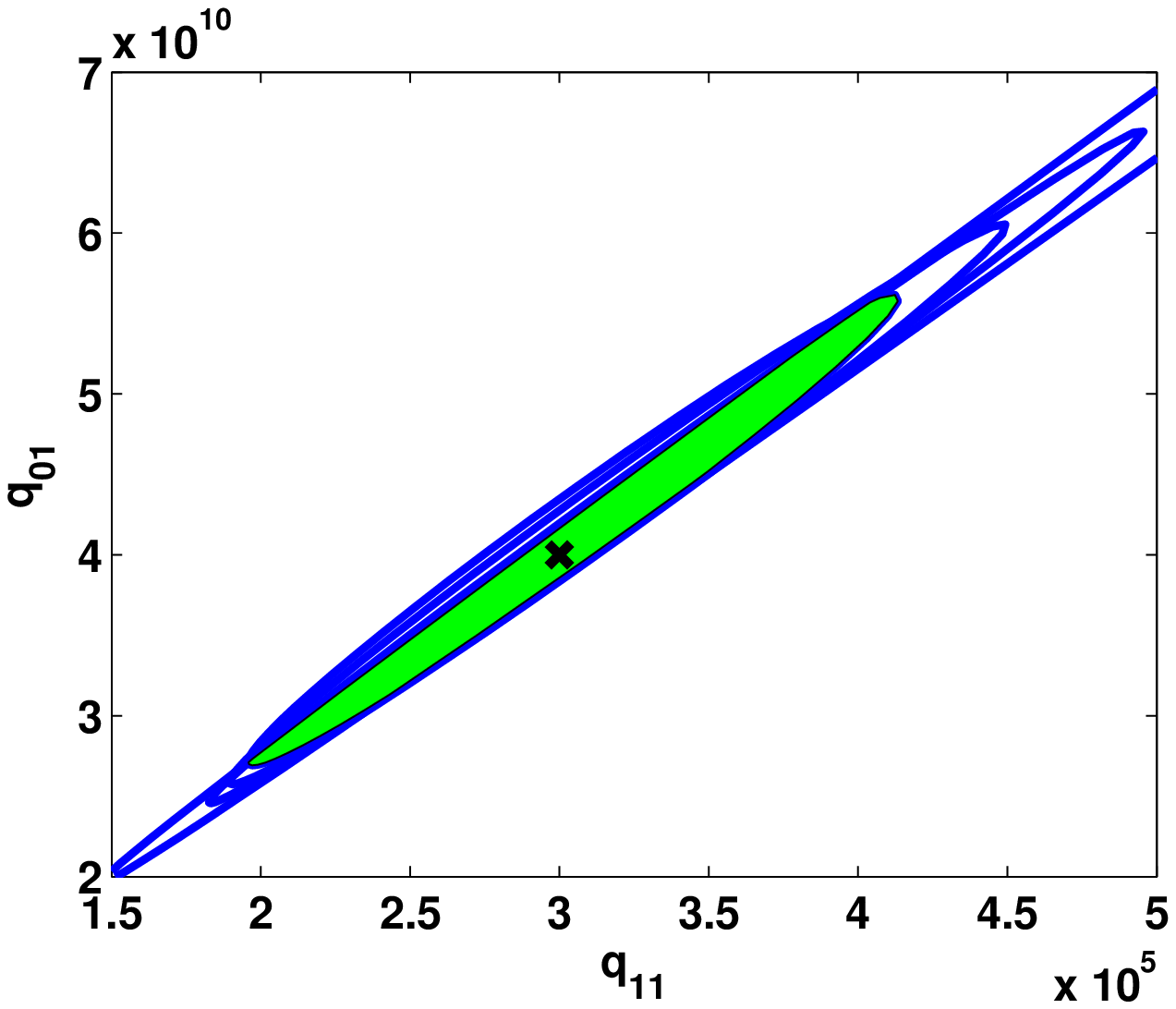}
                \caption{$\vert W_{S}S \vert <1$ and $\vert W_{T}T \vert <1$, $\forall\omega\in [0,\infty )$}
                \label{fig:gull}
        \end{subfigure}%
        \begin{subfigure}[b]{0.5\textwidth}
                \includegraphics[width=\textwidth]{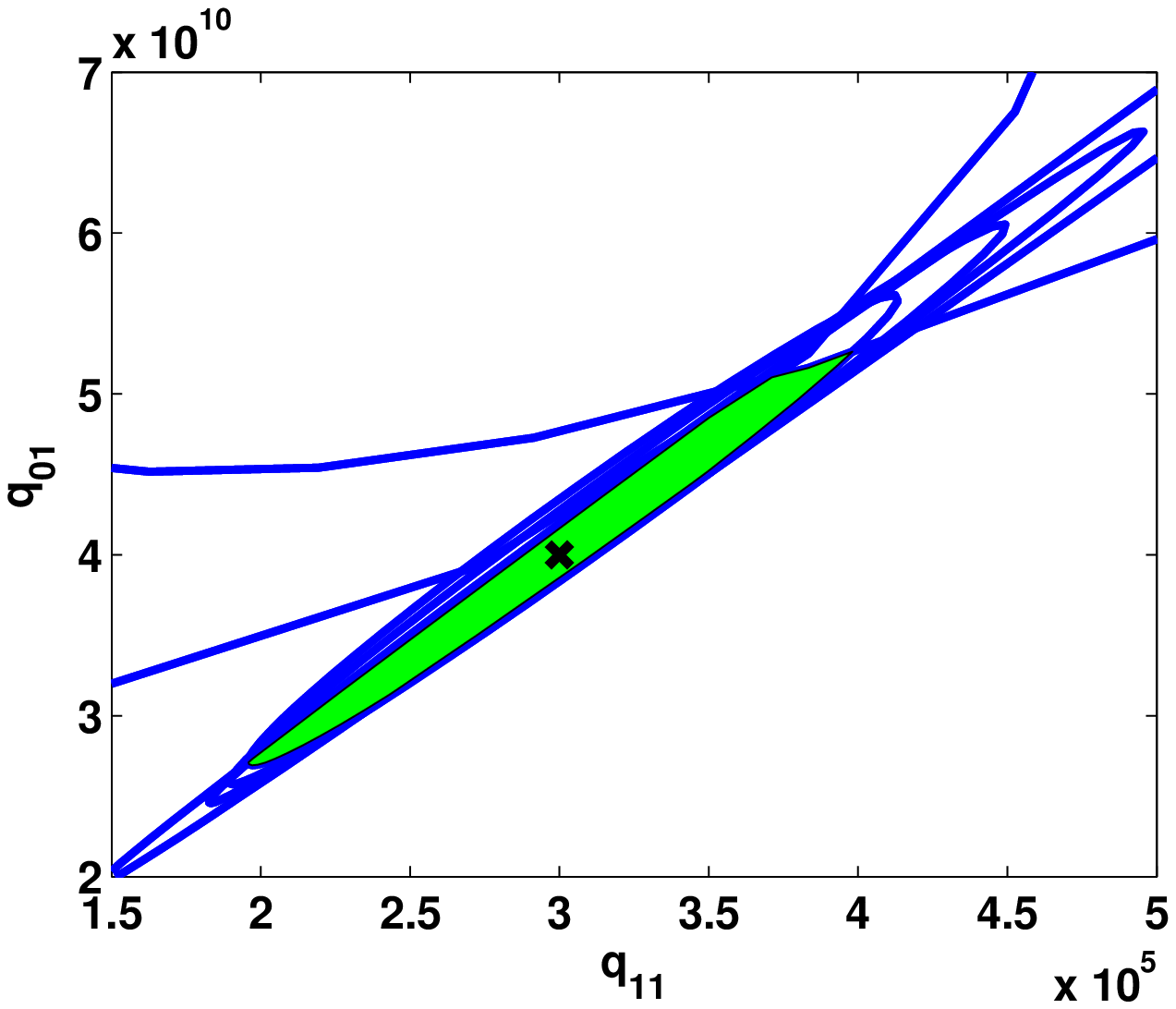}
                \caption{$\vert W_{S}S \vert + \vert W_{T}T \vert <1$, $\forall\omega\in [0,\infty )$}
                \label{fig:tiger}
        \end{subfigure}
        \caption{Parameter space region of the low-pass filter $q_{p}(s)$.}\label{fig:animals}
\end{figure}

\begin{figure}
        \centering
        \begin{subfigure}[b]{0.5\textwidth}
                \includegraphics[width=\textwidth]{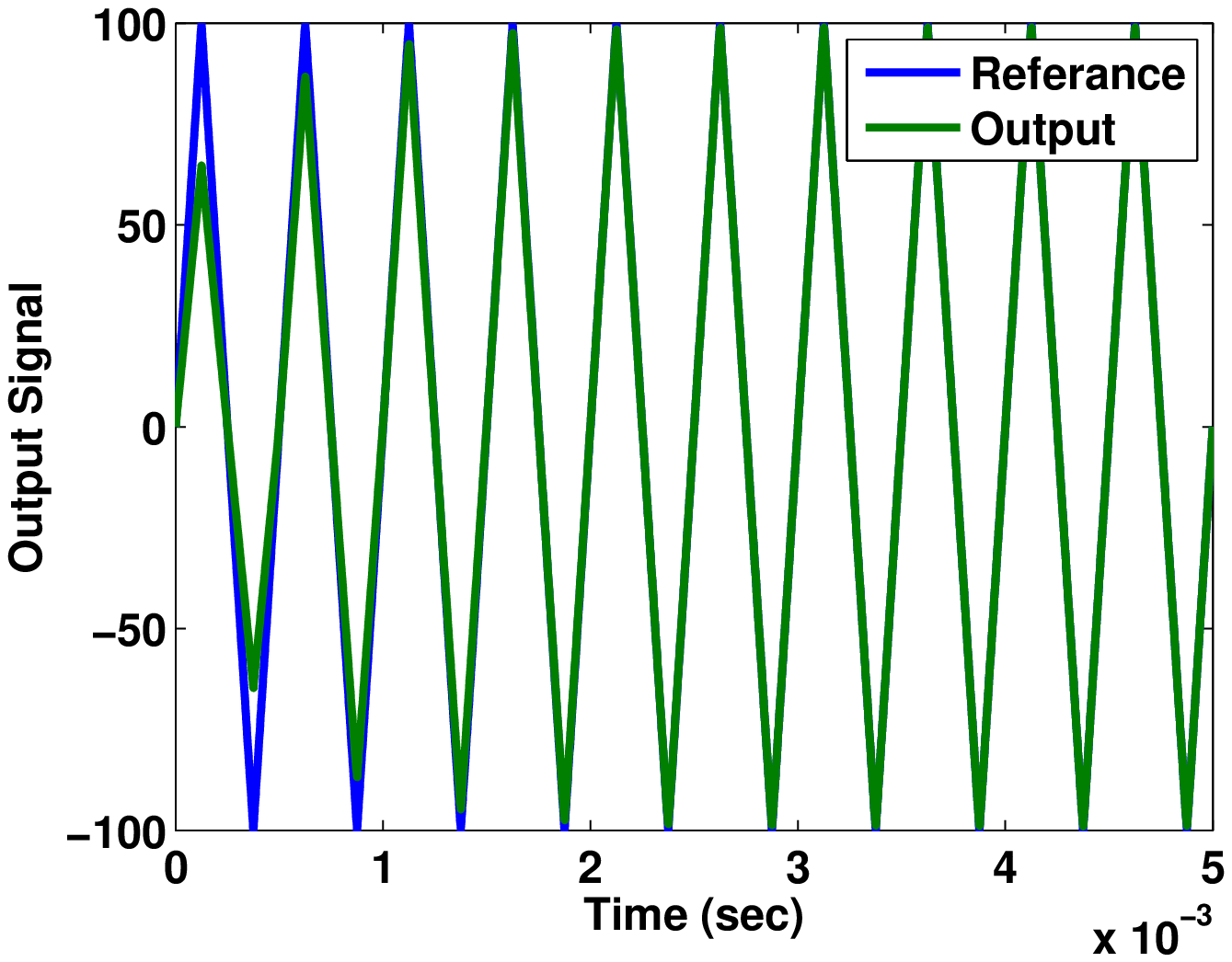}
                \caption{Output signal of the controlled system}
                \label{fig:gull}
        \end{subfigure}%
        \begin{subfigure}[b]{0.5\textwidth}
                \includegraphics[width=\textwidth]{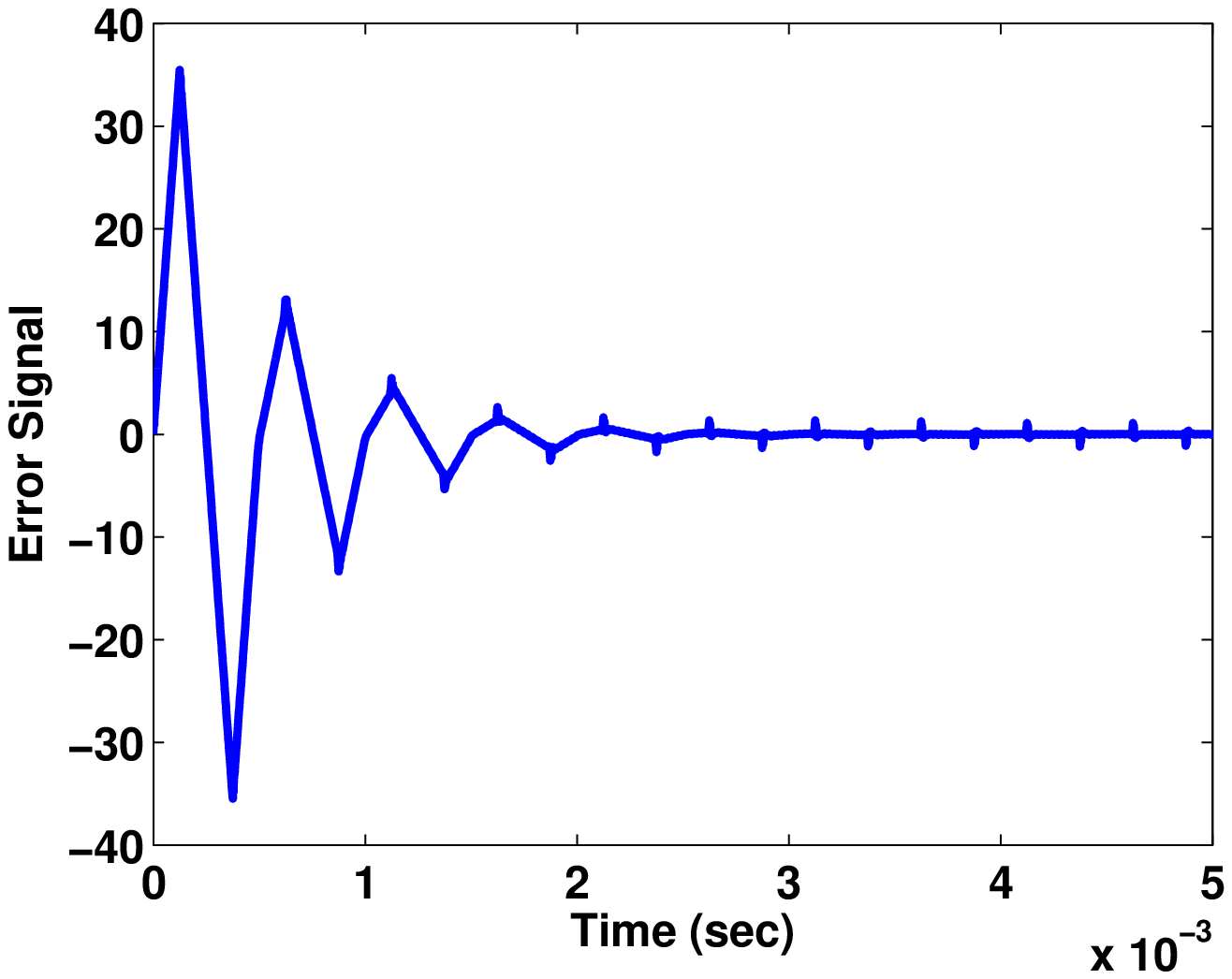}
                \caption{Error signal of the controlled system}
                \label{fig:tiger}
        \end{subfigure}
        \caption{Simulation results for triangular wave input at $2$ kHz.}\label{fig:animals}
\end{figure}

The sensitivity constraints are specified at a set of discrete frequencies. The periodic input command of the high speed AFM
scanner has a period of $\tau_{d}$ sec. The specific numerical values of the chosen weights used in the computation of the
controller parameters are seen in Table 2. The frequencies corresponding to the weights in Table 2 are shown with dots in the
Bode magnitude plot of Fig. 4. The overall region calculated for nominal performance, robust performance and stability robustness
can be seen in Fig. 5b. The method in Aksun G\"uven\c{c} and G\"uven\c{c}~\cite{GuG:06} where the solution region is obtained by
intersecting the nominal performance and stability robustness plots is presented in Fig. 5a for allowing easy comparison with an
existing method. Note that the sufficient stability condition $\mathcal{R}(\omega)<1$ for all $\omega\in [0,\infty )$ was also mapped and nominal stability is thus satisfied for the two solution regions shown in Fig. 5. Also note that the results in reference~\cite{GuG:06} are for the more conservative problem where nominal performance and robust stability solution regions are found separately. This conservatism
is reduced by the method in the present paper where the robust performance condition in Eq.~\eqref{eq:Robust_performance_stability} is handled directly. The method in this paper is more general in nature and contains the method of~\cite{GuG:06} as a special case.

The mathematical model of the high-speed AFM scanner cannot be fitted very well for frequencies above $160$ kHz. A uniform weight
for the robust stability requirement for frequencies above this value is chosen here as $\vert T \vert < 0.05$ for $f \geq 160$ kHz. This
corresponding discontinuous weight $W_{T}$ has been shown graphically with red-colored cross sign in Fig. 4. The relative
multiplicative error $\vert (G-G_{n})/G_{n} \vert$ has to be below the weight specified in the stability robustness considerations given
in Fig 4. The intersection of the regions, which are calculated in order for the nominal performance, the robust performance and the
robust stability requirements, in the $q_{01}-q_{11}$ controller parameter space is filled with green color. The designation
procedure is concluded by choosing a point in the controller parameter plane given in Fig. 5. The solution within this region
is chosen arbitrarily in this example and is point is marked with a cross in Fig. 5. The simulation result for a triangular wave
input with the period $2$ kHz and amplitude can be seen in Fig. 6. This result shows the effectiveness of the repetitive controller
in decreasing the steady state error while tracking a periodic input signal.

\section{Conclusion}

A multi-objective parameter space repetitive controller design procedure for satisfying a robust performance objective was
presented here. The main idea was to use a simple easily implementable structure for the repetitive controller filters and
compute solution regions in the chosen controller parameter space where frequency domain specifications on the nominal performance
at low frequencies ($W_{T}=0$), robust performance at intermediate frequencies and robust stability at high frequencies ($W_{S}=0$)
are satisfied. The abovementioned method is well suited to the structure of a repetitive control system with discrete frequencies
of interest and the computations were also quite fast. The proposed method is successfully applied to the infinite
dimensional nature of the repetitive control system with its inherent time delay. The effectiveness of the proposed method was
demonstrated by carrying out a design and simulation study for high speed AFM scanner position control problem.

\bibliographystyle{IEEEtran}
\bibliography{bibliography_IEEE}

\end{document}